\documentclass{amsart}

 \newtheorem{thm}{Theorem}[section]

 \theoremstyle{definition}
 \newtheorem{defn}[thm]{Definition}
 \theoremstyle{remark}

 \numberwithin{equation}{section}

\usepackage{bm}
\usepackage{graphicx}
\usepackage{amssymb}
\usepackage{hyperref}

\newcommand{\be}{\begin{equation}}
\newcommand{\ee}{\end{equation}}
\newcommand{\beq}{\begin{eqnarray}}
\newcommand{\eeq}{\end{eqnarray}}

\newcommand{\semmi}[1]{}

\def\be{\bigskip}
\def\beq{\bigskip}
\def\eeq{\bigskip}
\def\ee{\bigskip}
\def\bea{\begin{eqnarray}}
\def\eea{\end{eqnarray}}

\begin{document}

\title{Maximum Likelihood Estimation for Markov Chains}

\author{Iuliana Teodorescu}
\address{Department of Statistics, University of New Mexico, Albuquerque, NM 87131}

\begin{abstract}
A new approach for optimal estimation of Markov chains with sparse transition matrices is presented.
\end{abstract}

\maketitle

\tableofcontents

\section{Mathematical Framework}

We begin with a formal mathematical definition of a Markov chain:

\begin{defn}
\noindent Let $n$ and $d$ be elements of 
{\bf {N}}, such that $n \ge 1$ and $d \ge 1$. Define $\Omega = \{ 1, \ldots, d  \} $. Consider a sequence of random variables  $\{ X_{1}, X_{2}, \ldots, X_{n} \}$ such that 

\begin{equation}
P_{ij} = P(X_{k+1} = j | X_{k} = i)
\end{equation}

\noindent is independent of $k$ for all $i$ and $j$ in $\Omega$. Then the sequence $\{X_{1}, X_{2}, \ldots, X_{n} \}$ is a Markov chain with state space $\Omega$ and transition probabilities $P_{ij}$ for $i$ and $j$ in $\Omega$.

\end{defn}

\noindent It follows from this definition  that a Markov chain with known probability distribution of the initial state is completely characterized by a $d \times d$ matrix containing the transition probabilities $P_{ij}$,

$$ P = 
\left [ 
\begin{array}{cccc}
P_{11} & P_{12} & \ldots & P_{1d} \\
P_{21} & P_{22} & \ldots & P_{2d} \\
\vdots & \vdots & \ddots & \vdots \\
P_{d1} & P_{d2} & \ldots & P_{dd} \\
\end{array}
\right ]. $$

\noindent This matrix is called the $transition$ $probability$ $matrix$. Since the elements of row $i$ of this matrix represent the conditional probabilities for all possible state changes from state $i$, they must satisfy

\begin{equation}
\sum_{j=1}^{d}P_{ij} = 1,
\end{equation}

\noindent  for all $i \in \Omega$. For a Markov chain with known transition probability matrix, the most likely state as $n \rightarrow \infty$ can be calculated as follows. 
Define a vector $V_{k}$ so that the $i^{th}$ element of $V_{k}$ is the unconditional probability that the Markov chain is in state $i$ at time $k$. Hence, $(V_{k})_{i} = P(X_{k} = i),$ where  $V'_{k} = [(V_{k})_{1}, \ldots, (V_{k})_{d}].$ 

The probability $(V_{k+1})_{i}=P(X_{k+1}=i)$ can be related to the vector $V_{k}$ using the Law of Total Probability,
$$(V_{k+1})_{i}=P(X_{k+1}=i) = \sum_{j=1}^{d}P(X_{k}=j)P(X_{k+1}=i | X_{k} =j)=\sum_{j=1}^{d}P_{ji} \cdot (V_{k})_{j}.$$

\noindent Hence $V_{k+1}=P'V_{k}$. One can then use an inductive argument to
establish that $V_{k+1} = (P')^{k}V_{1}.$ Here $V_{1}$ is a vector of probabilities corresponding to the distribution of the initial state of the Markov chain. Hence
$$P(X_{1}=j)=(V_{1})_{j}$$

\noindent for $j=1, \ldots, d.$ 

The limiting, or steady state, probabilities, if they exist, are then given by 
\begin{equation}
\Pi^{(i)}=\lim_{n \rightarrow \infty} [(P')^{n}]\cdot V_{1}^{(i)}.
\end{equation}

\noindent Since $\displaystyle{[\Pi^{(i)}]_{j}= \sum_{k=1}^{d} \lim_{n \rightarrow \infty} [(P')^{n}]_{jk}\delta_{ik} = \lim_{n \rightarrow \infty} [(P')^{n}]_{ji}}$, it follows that 
$[\Pi^{(i)}]' = [\Pi^{(i)}_{1}, \ldots, \Pi^{(i)}_{d}]$ is the $i^{th}$ row of $P_{\pi} = \displaystyle{ \lim_{n \rightarrow \infty} P^{n}}$.

Under certain conditions [14], the limit will exist and the rows
of $P_{\pi}$ will be identical. We will denote one of these rows as $\Pi$. The elements of $\Pi$ correspond to the long-range probabilities that the Markov chain is in each of the states. In some instances $\Pi$ can be found analytically. 

\noindent {\bf {Example:}} Consider a Markov chain with transition probability matrix
$$
P=\left [
\begin{array}{cc}
\frac{1+a}{2} & \frac{1-a}{2} \\
\frac{1-a}{2} & \frac{1+a}{2} 
\end{array}
\right ],
$$

\noindent where $0 \le a < 1.$ A simple induction arguments shows that

$$
P^{n} =  \left [
\begin{array}{cc}
\frac{1+a^{n}}{2} & \frac{1-a^{n}}{2} \\
\frac{1-a^{n}}{2} & \frac{1+a^{n}}{2} 
\end{array}
\right ],
$$

\noindent for all integers $n \ge 1.$ Since $0 \le a < 1,$ $\displaystyle{\lim_{n \rightarrow \infty}a^{n} = 0  },$ so the limit $\displaystyle{P_{\pi} = \lim_{n \rightarrow \infty}P^{n}}$ exists and the rows of $P_{\pi}$ are identical:
$$
P_{\pi} =  \left [
\begin{array}{cc}
\frac{1}{2} & \frac{1}{2} \\
\frac{1}{2} & \frac{1}{2} 
\end{array}
\right ].
$$

%\newpage

\section{Estimation of the Transition Probability Matrix}

In most practical cases, the transition probability matrix is unknown and it must then
be estimated based on the observations. Let $X_{1}, X_{2}, \ldots, X_{n}$ be $n$ consecutive observations from a Markov chain.  The maximum likelihood estimator of the matrix $P$, which we will denote as $\widehat{P}$, is defined as follows [3]:

\begin{description}

\item [1) ] For each state $i \in \Omega$, let $n_{i}$ be the number of times that state $i$ is observed in $X_{1}, X_{2}, \ldots, X_{n-1}$.
\item [2) ] If $n_{i} = 0$ (the state is not represented in the chain, except maybe for the last position), 
then we formally define all probabilities of transition from the state $i$ to any state $j \ne i$ to be $0, \hat{P}_{ij} = 0,$ 
for every $j \ne i$. Therefore, by (2), we have $\widehat{P}_{ii} = 1$.
\item [3) ] If $n_{i} > 0$, let $n_{ij}$ be the number of observed consecutive transitions from state $i$ to state $j$ in 
$X_{1}, X_{2}, \ldots, X_{n}$. In this case, $\widehat{P}_{ij}=\frac{n_{ij}}{n_{i}}$, for $j=1,\ldots,d$. 

\end{description}

Note that the final observed state of the chain is not counted in Step 1 because we do not observe any transitions from this state. Hence, we only observe $n-1$ transitions. 
Note also that the estimate $\widehat{P}$ is a valid transition probability matrix.

Since the transition probability matrix has $d^{2}$ elements, it is
natural to rewrite $P$ as a column vector with $d^{2}$ elements [1]:
$$  P_{v} =vec(P) = \left [ 
\begin{array}{c}
P_{11} \\ P_{12} \\ \vdots \\ P_{1d} \\
 \vdots  \\
P_{d1} \\ \vdots \\ P_{dd} \\
\end{array} \right ] $$

\noindent which allows us to concentrate on properties of the
random vector $P_{v}$. This vector has $d^{2}$ elements, labeled by a two-digit index. For instance, $P_{ij}$ is the element found on the row $k = j + (i-1)d$ of the vector $vec(P): $ $P_{ij} = (P_{v})_{k}.$  

The properties of the maximum likelihood estimator $\widehat{P}$ have been
studied extensively [1]. In particular, $\widehat{P}$ can be shown to be asymptotically normal and consistent. The limiting probabilities computed from $\widehat{P}$ are also consistent estimates of the true limiting probabilities. These results are presented in the two theorems below.

Let $\widehat{P}_{n}$ be the maximum likelihood estimator corresponding to $n$ observations $X_{1}, \ldots, X_{n}$ from a Markov chain with transition probability  matrix $P$. Let $(\widehat{P}_{v})_{n}$ and $P_{v}$ be the vector forms of $\widehat{P}_{n}$ and $P$, respectively. The following theorem describes the asymptotic properties of the vector $(\widehat{P}_{v})_{n}$ as $n \rightarrow \infty.$

\begin{thm}
\noindent  As $n \rightarrow \infty,$

\begin{equation}
\sqrt{n} \left [ (\widehat{P}_{v})_{n} - P_{v} \right ] \stackrel{w}{\rightarrow} N(O,\Sigma_{P}),  
\end{equation}

\noindent where $\Sigma_{P}$ is given by

\begin{equation}
(\Sigma_{P})_{(ij,kl)} = \delta_{ik}P_{ij}(\delta_{jl}-P_{il}).
\end{equation}

\end{thm}

\noindent Here, $\Sigma_{P}$ is a square $d^{2} \times d^{2}$ matrix.  The matrix element displayed corresponds to the row $j+(i-1)d$ and the column $l + (k-1)d$. 

Now assume that for all integers $n > 0,$ the limit $\displaystyle{\lim_{m \rightarrow \infty}[\widehat{P}_{n}]^{m}}$ exists and has all rows identical. Denote by $\widehat{\Pi}_{n}$ and $\Pi$  the steady-state probabilities corresponding to $\widehat{P}_{n}$ and $P$, respectively. The following theorem establishes the consistency of the estimates of steady-state probabilities.

\begin{thm}
For all $i$, $(\widehat{\Pi}_{n})_{i} \rightarrow (\Pi)_{i}$,  with probability 1,  as $n \rightarrow \infty,$ where $(\widehat{\Pi}_{n})_{i}$ and $(\Pi)_{i}$ are the $i^{th}$ elements of $\widehat{\Pi}_{n}$ and $\Pi$ respectively.
\end{thm}

These results provide an asymptotic justification of the use of $\widehat{P}$ to estimate $P$. When the sample size is not sufficiently large, the asymptotic results given in previous results may not hold. In these cases, the bootstrap method, which is outlined in the next section, can be used to find approximate results corresponding to those given above.

\section{The Bootstrap Method}

Let $X$ be a random variable with distribution function $F$ and let ${\bf{X}} = (x_{1}, \ldots, x_{n})'$ be an observed sample from $F$. Suppose $R({\bf {X}},F)$ is a statistical quantity that depends in general on both  the unknown distribution $F$ and on the sample ${\bf {X}}$. For example, $R({\bf {X}}, F)$ could be an estimator of an unknown parameter. If $F$ is unknown, then the exact distribution of the random variable $R({\bf {X}},F)$ is generally unknown. 

In 1979, Efron [5] proposed the $bootstrap$ method to
nonparametrically estimate the distribution of  $R({\bf {X}},F)$. The method consists of the following three steps:

\begin{description}
  \item[(i)] From the observed sample $\bf{X}$, use the empirical distribution function,
$\widehat{F}_{n}$, as an estimate of the probability function $F$. The empirical distribution function is defined by $\widehat{F}_{n}(x) = \frac{n(x)}{n},$ where $n(x)$ is the number of values $x_{i}$ in {\bf {X}} that are less than or equal to $x$. 
  \item[(ii)] Draw $B$ samples of size $n$ from $\widehat{F}_{n}$ conditional on {\bf {X}}. Denote these as ${\bf{X}}^{*}_{j}$, for $j = 1, \ldots , B.$
  \item[(iii)] For each sample ${\bf{X}}^{*}_{j}$, compute $R^{*}_{j} = R({\bf {X}}^{*}_{j},\widehat{F}_{n})$ and 
approximate the distribution of $R({\bf {X}},F)$ with the empirical distribution of  $R^{*}_{1}, \ldots, R^{*}_{B}$.
\end{description}

The samples ${\bf {X}}^{*}_{1}, \ldots, {\bf {X}}^{*}_{B}$ are called resamples and the empirical distribution of $R^{*}_{1}, \ldots, R^{*}_{B}$ is called the bootstrap estimate of the distribution of $R$, or simply the bootstrap distribution of $R^{*}$.

The $bootstrap$ $principle$ states that the empirical distribution of
$R^{*}_{1}, \ldots, R^{*}_{B}$ is a good approximation to the true
distribution of $R({\bf {X}}, F)$. Several authors have proven that the approximation is asymptotically valid for a large number of statistics of interest, and underlying populations, under some regularity conditions. See [4] and [6].

In [11], Kulperger and Prakasa Rao studied the applicability of
the bootstrap method to the problem of estimating properties of Markov chains. Working under certain assumptions, they proved the following Central Limit Theorem for the bootstrap maximum likelihood estimator matrices.

Let $X_{1}, \ldots , X_{n}$ be $n$ observations from a Markov chain with transition probability matrix $P$ and let $\widehat{P}_{n}$ be the maximum likelihood estimator of $P$ computed on the sample. Generate a bootstrap chain, $X^{*}_{1}, \ldots , X^{*}_{n}$, by generating a Markov chain with  transition probability matrix $\widehat{P}_{n}$, conditional on $X_{1}, \ldots , X_{n}$. Denote the maximum likelihood estimator for the bootstrap chain by $\widehat{P}^{*}_{n}$. Let $(\widehat{P}_{v}^{*})_{n}$ and $(\widehat{P}_{v})_{n}$ be the vector forms of $\widehat{P}_{n}^{*}$ and $\widehat{P}_{n}$, respectively.

\begin{thm}
There is a sequence $N_{n} \in {\mathbb {N}}$, such that 
\begin{equation}
\sqrt{N_{n}}\left [(\widehat{P}^{*}_{v})_{n} - (\widehat{P}_{v})_{n} \right ] \stackrel{w}{\rightarrow} N(0,\Sigma_{P}),
\end{equation}
\noindent as $n \rightarrow \infty$ and $N_{n} \rightarrow \infty$, where 
\begin{equation}
(\Sigma_{P})_{(ij,kl)} = \delta_{ik}P_{ij}(\delta_{jl}-P_{il}).
\end{equation}

\end{thm}

\noindent This result  indicates that the distribution of the bootstrap maximum likelihood estimator has similar asymptotic behavior as the distribution of the maximum likelihood estimator.

\section{The Bootstrap Method for Finite State Markov Chains}

When applied to the problem of estimating Markov chains, the bootstrap method consists of computing $\widehat{P}$ from the original chain, and then generating $B$ additional samples based on $\widehat{P}$. A uniform probability distribution for the initial state is used. For each of these resamples, a maximum likelihood estimator $\widehat{P}^{*}_{i}$, $i=1, \ldots, B$ is computed. Based on the vector sample $(\widehat{P}_{v}^{*})_{1}, \ldots, (\widehat{P}_{v}^{*})_{B},$  estimators for $E(P_{v})$ and $Cov(P_{v})$ can  be computed as follows:
$$\widehat{E(P_{v})} = \frac{1}{B}\sum_{k=1}^{B}(\widehat{P}^{*}_{v})_{k}, $$
$$\widehat{Cov(P_{v})} =  \frac{1}{B-1}\sum_{k=1}^{B} \left [ (\widehat{P}^{*}_{v})_{k}-\widehat{E(P_{v})}\right ] \cdot \left [ (\widehat{P}^{*}_{v})_{k}-\widehat{E(P_{v})} \right ]' ,$$

\noindent where $\widehat{Cov(P_{v})}$ is a square matrix of dimension $d^{2} \times d^{2}$.
 
The empirical distribution function for each element $(P_{v})_{ij}$ of the
vector $P_{v}$ can also be computed, based on the sample $[(\widehat{P}_{v}^{*})_{1}]_{ij}, \ldots, [(\widehat{P}_{v}^{*})_{B}]_{ij}.$ Denote this function by $\widehat{F}_{ij}.$ A $(1-\alpha)100\%$ confidence interval based on the percentile method of Efron (1979) (see also Reference [7]) for the element $(P_{v})_{ij}$ is given by $[\widehat{F}_{ij}^{-1}(\alpha),  \widehat{F}_{ij}^{-1}(1-\alpha)]$. Here, $x_{L} = [\widehat{F}_{ij}]^{-1}(\alpha)$ is the largest value of $x$ such that the number of elements in the sample $[(\widehat{P}_{v}^{*})_{1}]_{ij}, \ldots, [(\widehat{P}_{v}^{*})_{B}]_{ij}$ that are less than $x$ is smaller than $\alpha n$. Likewise, $x_{U} = [\widehat{F}_{ij}]^{-1}(1-\alpha)$ is the smallest value of $x$ such that the number of elements in the sample $[(\widehat{P}_{v}^{*})_{1}]_{ij}, \ldots, [(\widehat{P}_{v}^{*})_{B}]_{ij}$ that are smaller than $x$ is larger than $(1-\alpha) n$. Specifically, 
$$x_{L} = max \left \{x: (\widehat{F}_{n})_{ij}(x) \le \alpha  \right \}, {\hspace{0.2in}} x_{U} = min \left \{x: (\widehat{F}_{n})_{ij}(x) \ge 1-\alpha  \right \} .$$

The bootstrap procedure may not perform well in some circumstances. For example, under certain conditions, the matrix $\widehat{P}$ may not have a structure that is close to that of $P$.  To illustrate one of these situations, we consider the following numerical example.

\noindent {\bf {Example:}} Let the true transition probability matrix of a Markov chain be

\begin{equation}
P=\left [ 
\begin{array}{cccc}
0.25 & 0.25 & 0.25 & 0.25 \\
0.10 & 0.20 & 0.20 & 0.50 \\
0.05 & 0.10 & 0.10 & 0.75  \\
0.10 & 0.20 & 0.30 & 0.40  
\end{array} 
\right ]. 
\end{equation} 

\noindent Using the C code listed in Appendix A, we generated samples of length $n=10$ from this transition matrix, using an initial distribution of $V_{1} = (0.25, 0.25, 0.25, 0.25)'$. Ten such samples are listed in Table 1.

\begin{table}
  \centering
  \caption{The ten samples generated using the transition matrix in (8)}\label{}

\begin{tabular}{cc}
\\
\hline Sample Number &  Generated Sample \\ \hline
  1 &   3, 4, 2, 4, 3, 4, 3, 4, 4, 1 \\
  2 &   2, 2, 1, 4, 4, 4, 1, 1, 4, 3 \\
  3 &   3, 2, 4, 3, 4, 2, 2, 4, 3, 4 \\
  4 &   2, 4, 4, 4, 2, 4, 4, 2, 4, 3 \\
  5 &   3, 2, 2, 4, 3, 4, 4, 4, 3, 4 \\
  6 &   4, 4, 3, 4, 3, 4, 4, 3, 4, 4 \\
  7 &   2, 2, 4, 4, 2, 4, 2, 3, 4, 4 \\
  8 &   2, 3, 4, 3, 3, 3, 4, 1, 4, 2 \\
  9 &   2, 4, 4, 1, 2, 3, 4, 4, 2, 3 \\
 10 &   1, 1, 4, 4, 1, 3, 4, 4, 4, 4 \\
\hline
\end{tabular}

\end{table}

%\skip

The first sample leads to the following maximum likelihood estimator $\widehat{P}$:

$$
\widehat{P}=\left [ 
\begin{array}{cccc}
1.00 & 0.00 & 0.00 & 0.00 \\
0.00 & 0.00 & 0.00 & 1.00 \\
0.00 & 0.00 & 0.00 & 1.00 \\
0.20 & 0.20 & 0.40 & 0.20 \\
\end{array} 
\right ] $$ 

\noindent Note that the estimate $\widehat{P}$ is significantly different from the original matrix $P$. The main 
difference is that $\widehat{P}$  is $sparse$ (has many null entries), while $P$ is not. Therefore, many
valid transitions will never occur in resamples based on the matrix $\widehat{P}$. Regardless of how many
bootstrap resamples we use, the fact that all the bootstrap maximum likelihood estimators $\widehat{P}^{*}$ are
sparse may cause the bootstrap method to give unreliable results.

Computing maximum likelihood estimators from the other samples generated from $P$ leads again to sparse estimators, 
though they may differ from the one listed above. This is because the sample size chosen is relatively small
compared to the total number of possible transitions ($n=10$, for $d^{2}=16$).  A maximum of only 60\% of all transitions
will be found in a given sample.

Another situation that leads to sparse estimators occurs when the matrix $P$ has elements with small probabilities.  In this case, it is the existence of
$rare$ transitions (corresponding to the small probabilities) that causes the problem. For instance, if we use the matrix $\widehat{P}$
as the true $P$ matrix, we obtain the samples listed in Table 2.

%\newpage

\begin{table}
  \centering
  \caption{The ten samples generated using $\widehat{P}$}\label{}

\begin{tabular}{cc}
\\
\hline Sample Number &  Generated Sample \\ \hline
  1 &  3, 4, 1, 1, 1, 1, 1, 1, 1, 1 \\
  2 &  2, 4, 1, 1, 1, 1, 1, 1, 1, 1 \\
  3 &  3, 4, 3, 4, 3, 4, 1, 1, 1, 1 \\
  4 &  2, 4, 3, 4, 1, 1, 1, 1, 1, 1 \\
  5 &  3, 4, 1, 1, 1, 1, 1, 1, 1, 1 \\
  6 &  4, 4, 2, 4, 3, 4, 3, 4, 2, 4 \\
  7 &  2, 4, 4, 4, 1, 1, 1, 1, 1, 1 \\
  8 &  2, 4, 4, 3, 4, 2, 4, 1, 1, 1 \\
  9 &  2, 4, 4, 1, 1, 1, 1, 1, 1, 1 \\
 10 &  1, 1, 1, 1, 1, 1, 1, 1, 1, 1 \\
\hline
\end{tabular}

\end{table}

\bigskip

\noindent The  maximum likelihood estimator of the first sample is:
$$\widehat{P}=\left [ 
\begin{array}{cccc}
1.00 & 0.00 & 0.00 & 0.00 \\
0.00 & 1.00 & 0.00 & 0.00 \\
0.00 & 0.00 & 0.00 & 1.00 \\
1.00 & 0.00 & 0.00 & 0.00 \\

\end{array} 
\right ]. $$ 

\noindent As indicated earlier, increasing the number of samples does not help, since all the estimators will be sparse. To
avoid this from happening, one should use a non-sparse matrix to generate the resamples. 

Next, we will describe a way of solving this problem, by $smoothing$ the maximum likelihood estimators. This  procedure replaces a sparse estimator by a modified version where  all of the entries are positive.

\section{Smoothed Estimators}

As indicated in the previous section, a problem related to 
estimating the transition probability matrix from observed sample chains
is the possibility that
some states of the system are too rare to occur in a limited experiment. A similar
result is obtained when the chain length, $n$, is small compared to the
total number
of possible transitions, $d^{2}$. In this case only a fraction of all the possible 
transitions will be present in any given sample. 
When this happens, a particular transition may not be observed in the sample,
even though the probability of this transition occuring is greater than 0.

When a sparse estimator $\widehat{P}$ is obtained from the initial chain, the
impact on the bootstrap method is significant. If we assume that $\widehat{P}_{ij}=0$ for some $i$ and $j$, then
a transition from state $i$ to state $j$ will never be observed in any of the
resamples, even though it may be possible in the actual Markov chain. A
similar problem occurs in the case of using the bootstrap on independent
discrete data. In [9] and [13], the authors exhibit several
examples where sparse data causes the bootstrap to perform poorly. 

One solution to this problem is to increase the sample size. When a larger
sample size is not feasible, the following method can be used. Since the
cause of the problem is the fact that $\widehat{P}$ is sparse, we can
attempt to generate the bootstrap resamples based on a slightly different
matrix, whose entries are all positive. We call this matrix the $smoothed$
version of $\widehat{P}$ and denote it by $\widetilde{P}$. It is given by

\begin{equation}
\widetilde{P}_{ij} = \frac{1}{\omega} [\widehat{P}_{ij} + n^{-u}],
\end{equation}

\noindent where 
$$\omega = \sum_{j=1}^{d}[\widehat{P}_{ij} + n^{-u}] = \sum_{j=1}^{d} \widehat{P}_{ij} +\sum_{j=1}^{d}n^{-u} = 1 + n^{-u}d,  $$

\noindent and $u > 0$ is a positive smoothing parameter. 

The form of this smoothed matrix is based on simple smoothers that are
used for multinomial distributions. See, for example,  [8] and
[16]. Note that from the definition, we obtain 

$$ \sum_{j=1}^{d}\widetilde{P}_{ij} = \frac{1+dn^{-u}}{\omega} = 1, {\hspace{0.1in}} {\textrm{ for all }} {\hspace{0.1in}} i=1, \ldots, d,$$

\noindent so that $\widetilde{P}$ is a valid transition probability matrix.

The choice of the smoothing parameter $u$ presents some difficulty. It is technically possible to specify a performance criterion for $\tilde{P}$ in terms of some measure of the performance of the resulting bootstrapping method. The parameter $u$ could then be chosen to optimize this criterion. However, it is unlikely that such a method would be feasible in practice, and is well beyond the scope of this study. Nevertheless, we will justify some general properties that $u$ should follow. These will ensure that the smoothing does not asymptotically affect the behavior of the generated Markov chains.

The criterion we choose is to select the smoothing parameter such that
$\widetilde{P}$ is a consistent estimator of $P$ at the same rate as
$\widehat{P}$.

%\newpage

\section{Asymptotic Properties of Smoothed Estimators}
 
In the following, we consider $n$ observations $X_{1}, X_{2}, \ldots, X_{n}$ from a Markov chain and establish the asymptotic properties of the smoothed  estimator of the transition probability matrix. We begin by proving some general properties.

In order to study the asymptotic properties of estimators, we must introduce the following equivalence relation for matrices. 

\bigskip
Let $\{ P_{n} \}$ and $\{ R_{n} \}$ be two sequences of $d \times d$ matrices, for $n=1, 2, \ldots$. Suppose there is an $r > 0$ such that the sequence $n^{r}(E_{n})_{ij} = n^{r}(P_{n}-R_{n})_{ij}$ has the property that it remains bounded as $n \rightarrow \infty$ for all $i, j = 1, \ldots, d$. Then as $n \rightarrow \infty$ 

\begin{equation}
P_{n} = R_{n} + O(n^{-r}).
\end{equation}
\bigskip

\noindent Here, $O(n^{-r})$ represents any sequence of matrices properly bounded. 

The following theorem describes the asymptotic consistency property of the smoothed estimator defined earlier.

\begin{thm}
Suppose $\widehat{P} = P + O(n^{-k})$ as $n \rightarrow \infty$ for some $k > 0$. Then $\widetilde{P} = P +O(n^{-k})$ as $n \rightarrow \infty$ as long as $u \ge k$.
\end{thm}

%\skip

\noindent  {\bf {Proof:}}

%\skip 

\noindent Consider the function $f(x) = (1+x)^{-1}$. A Taylor expansion of $f$ around $x = 0$ is 
$$f(x) = 1 + O(x) {\textrm { as }} x \rightarrow 0. $$
We can rewrite $\omega^{-1} = f(n^{-u}d)$ so that
\begin{equation}
\omega^{-1} = 1 + O(n^{-u}), {\textrm { as }} n \rightarrow \infty, 
\end{equation}
\noindent since $n^{-u}d$ remains bounded as $n \rightarrow \infty$ for fixed integer $d \ge 1$.

\noindent Now computing 
$$n^{u}[n^{-u}\omega^{-1}] = \omega^{-1} = 1 + O(n^{-u}), $$
 
\noindent which by definition [15] remains bounded as $n \rightarrow \infty$, so

\begin{equation} 
n^{-u}\omega^{-1} = O(n^{-u}).
\end{equation}

In matrix notation, this result can be rewritten as

\begin{equation}
\widetilde{P} = \omega^{-1}\widehat{P} + n^{-u}\omega^{-1}J,
\end{equation}

\noindent where $J$ is a $d \times d$ matrix with all entries equal to 1. We conclude that:

\begin{equation} 
\widetilde{P} = \widehat{P} + a_{n}\widehat{P} + b_{n}J,
\end{equation}

\noindent where the sequences $\displaystyle{n^{u}a_{n}}$ and
$\displaystyle{n^{u}b_{n}}$ remain bounded as $n \rightarrow \infty$.  Then for all $i, j = 1, \ldots, d$, $0 \le \widehat{P}_{ij} \le 1$ and $J_{ij} = 1$, so  $\displaystyle{n^{u}[a_{n}\widehat{P}_{ij} + b_{n}J_{ij}]}$ remains bounded as $n \rightarrow \infty$. Therefore,

\begin{equation}
\widetilde{P} = \widehat{P} + O(n^{-u}), {\hspace{0.25in}} {\textrm{as }}
n \rightarrow \infty.
\end{equation}

Since $\widehat{P} = P + O(n^{-k}),$ we can write

\begin{equation}
\widetilde{P} = P + A_{n} + B_{n}, 
\end{equation}

\noindent where $\displaystyle{ n^{k}(A_{n})_{ij} }$ 
and $\displaystyle{ n^{u}(B_{n})_{ij}}$ remain bounded as $n \rightarrow \infty$. Then for all $k \le u,$ $\displaystyle{[ n^{k}(A_{n})_{ij} + n^{k-u}n^{u}(B_{n})_{ij} ]}$ remains bounded as $n \rightarrow \infty$, so,

\begin{equation}
\widetilde{P} = P + O(n^{-k}), 
\end{equation}

\noindent as long as $k \le u$.

As shown in [1] and [3], the exponent $k$ is
usually equal to 0.5. Therefore, any choice of $u$ such that $u \ge 0.5$ will ensure that $\widetilde{P}_{n}$ preserves the asymptotic consistency property of $\widehat{P}_{n}$. 

\section{Performance of Smoothed Estimators}

To compare the performance of the smoothed and unsmoothed estimators, we present two Examples.

 {\bf {Example:}} In this example we explore the behavior of the bootstrap bias estimator using $\widehat{P}$ and $\widetilde{P}$. We use the transition probability matrix from the example given in (8). The true probability matrix is:

$$P=\left [ 
\begin{array}{cccc}
0.25 & 0.25 & 0.25 & 0.25 \\
0.10 & 0.20 & 0.20 & 0.50 \\
0.05 & 0.10 & 0.10 & 0.75  \\
0.10 & 0.20 & 0.30 & 0.40  \\
\end{array} 
\right ]. $$ 

\noindent Using a chain generated from $P$, with uniform distribution probability for the initial state, the following maximum likelihood estimator is computed:

$$\widehat{P}=\left [ 
\begin{array}{cccc}
0.111111 & 0.222222 & 0.222222 & 0.444444 \\
0.142857 & 0.142857 & 0.357143 & 0.357143 \\
0.000000 & 0.037037 & 0.185185 & 0.777778 \\
0.122449 & 0.183673 & 0.285714 & 0.408163 \\ 
\end{array} 
\right ]. $$ 

With smoothing parameter $u=0.5$ the smoothed maximum likelihood estimator is

$$\widetilde{P}=\left [ 
\begin{array}{cccc}
0.150794 & 0.230159 & 0.230159 & 0.388889 \\
0.173469 & 0.173469 & 0.326531 & 0.326531 \\
0.071429 & 0.097884 & 0.203704 & 0.626984 \\
0.158892 & 0.202624 & 0.275510 & 0.362974 \\
\end{array} 
\right ]. $$ 

\noindent After applying the bootstrap method, with $B=1000,$ the average estimator computed from the samples based on the unsmoothed matrix is found to be

$$\overline{\widehat{P}}=\left [ 
\begin{array}{cccc}
0.099799 & 0.220662 & 0.233852 & 0.445688 \\
0.145533 & 0.139222 & 0.359566 & 0.355678 \\
0.000000 & 0.035484 & 0.177676 & 0.786840 \\
0.121927 & 0.180267 & 0.287282 & 0.410525 \\ 
\end{array} 
\right ] $$ 

\noindent The average computed from the sample based on the smoothed estimator is given by

$$\overline{\widetilde{P}}=\left [ 
\begin{array}{cccc}
0.171886 & 0.240630 & 0.241419 & 0.346065 \\
0.196326 & 0.191570 & 0.307840 & 0.304264 \\
0.122615 & 0.141866 & 0.212862 & 0.522657 \\
0.183719 & 0.214623 & 0.270342 & 0.331316 \\  
\end{array}
\right ]. $$ 

As we can see, the smoothed estimator contains some information about the low-probability transitions of the system, while the standard maximum likelihood estimator does not. In particular, the element corresponding to the transition $3 \rightarrow 1$, which has the lowest probability for this chain, is strictly zero in the average maximum likelihood estimator, but not in the smoothed version. Since the average is computed from non-negative numbers, it follows that $(\widehat{P}^{*})_{31} = 0$ for all the resamples based on $\widehat{P}$. The bootstrap method based on $\widehat{P}$ leads to the conclusion that the transition $3 \rightarrow 1$ is not allowed in this chain. 

The bootstrap method based on $\widetilde{P}$ does not lead to the same conclusion, as all the elements of $\overline{\widetilde{P}}$ are positive. While $(\overline{\widetilde{P}})_{31}$ is not very close to the true value 0.05, the confidence interval for this element predicted by the bootstrap method based on $\widetilde{P}$ may have good coverage properties. The same conclusion holds for other statistical inference quantities. A simulation study of the coverage properties of the bootstrap confidence intervals is presented in the next section.

\noindent {\bf {Example:}} In this example we explore the asymptotic behavior of $\widehat{P}_{n}$ and $\widetilde{P}_{n}$ as $n \rightarrow \infty$. The matrix given in (8) is the true transition probability matrix of the system. Single samples of size 50, 100, 500, 1000 and 10,000 were generated based on $P$. For each sample, the estimators $\widehat{P}_{n}$ and $\widetilde{P}_{n}$ were computed. The matrices $\sqrt{n}(\widehat{P}_{n} - P)$ and $\sqrt{n}(\widetilde{P}_{n} - P)$ were then calculated. The results are listed in Tables 3 and 4.

\begin{table}
  \centering
  \caption{Asymptotic Behavior of $\sqrt{n}(\widehat{P}_{n} - P)$}\label{}

\begin{tabular}{cc}
\\
\hline Sample Size &  $\sqrt{n}(\widehat{P}_{n} - P)$ \\ \hline
$n = 50$ &  $\displaystyle{\left  .
\begin{array}{cccc}
 -0.353553 & -0.353553 & -0.353553 & 1.060660 \\
0.303046 & 0.606092 & -1.414214 & 0.505076 \\
-0.353553 & -0.707107 & 0.380750 & 0.679910 \\
0.176777 & -0.235702 & 0.530330  & -0.471404 \\
\end{array}
\right .}$ \\ \hline

 $n = 100$ & $\displaystyle{\left  .
\begin{array}{cccc}
-1.388889 & -0.277778 & -0.277778 & 1.944444 \\
0.428571 & -0.571429 & 1.571429 & -1.428571 \\
-0.500000 & -0.629630 & 0.851852 & 0.277778 \\
0.224490 & -0.163265 & -0.142857 & 0.081633 \\
\end{array}
\right . }$ \\ \hline

 $n = 500$ &  $\displaystyle{\left  .
\begin{array}{cccc}
-0.356819 & 0.118940  & 0.118940 & 0.118940 \\
1.000346  & -1.529941 & 2.000692 & -1.471097 \\
-0.396722 & -0.252459 & -0.432787 &  1.081969 \\
-0.372678 & -0.656623 & 0.301691 & 0.727609 \\
\end{array}
\right . }$     \\ \hline

 $n = 1000$ &   $\displaystyle{\left  .
\begin{array}{cccc}  
-1.956855 & -0.704468 & 1.800307 & 0.861016 \\
0.866102  & -1.893338 & 0.926527 & 0.100710 \\
-0.119582 & 0.026574  &  -0.637770 & 0.730779 \\
0.044008  & -0.792141 & 0.006286 & 0.741846 \\
\end{array}
\right . }$  \\ \hline

 $n = 10,000$ & $\displaystyle{\left  .
\begin{array}{cccc}  
-3.185526 & 1.503569 & 0.891948 & 0.790009 \\
0.154021 & -0.091273 & 0.992584 & -1.055333 \\
-0.764507 & -0.681914 & 0.758153 &  0.688267 \\
-0.028548 & -1.158239 & 0.261009 & 0.925773 \\
\end{array}
\right . }$  \\

\hline
\end{tabular}

\end{table}

%\skip

\begin{table}
  \centering
  \caption{Asymptotic Behavior of $\sqrt{n}(\widetilde{P}_{n} - P)$}\label{}

\begin{tabular}{cc}
\\
\hline Sample Size &  $\sqrt{n}(\widetilde{P}_{n} - P)$ \\ \hline
 $n = 50$ &  $\displaystyle{\left  .
\begin{array}{cccc}
-0.225814 & -0.225814 & -0.225814 & 0.677441 \\
0.576773 & 0.514849 & -0.775516 & -0.316107 \\
0.285145 & -0.068409 & 0.626403 & -0.843139 \\
0.496126 & -0.022803 & 0.210981 & -0.684304  \\
\end{array}
\right . }$ \\ \hline

 $n = 100$ & $\displaystyle{\left  .
\begin{array}{cccc}
-0.992063 &  -0.198413 & -0.198413 & 1.388889 \\
0.734694 & -0.265306 & 1.265306 & -1.734694 \\
0.214286 & -0.021164 & 1.037037 & -1.230159 \\
0.588921 & 0.026239 & -0.244898 & -0.370262  \\ 
\end{array}
\right .}$ \\ \hline

 $n = 500$ &  $\displaystyle{\left  .
\begin{array}{cccc}
-0.302675 & 0.100892 & 0.100892 & 0.100892  \\
1.357508 & -1.128134 & 1.866757 & -2.096130 \\
0.342084 & 0.294804 & 0.141840 & -0.778728  \\
0.192828 & -0.387335 & 0.086260 & 0.108245  \\
\end{array}
\right . }$     \\ \hline

 $n = 1000$ &   $\displaystyle{\left . 
\begin{array}{cccc}  
-1.737124 & -0.625364 & 1.598154 & 0.764335 \\
1.301476 & -1.503197 & 1.000032 & -0.798310 \\
0.604016 & 0.556217 & -0.033529 & -1.126703 \\
0.571694 & -0.525651 & -0.171962 & 0.125919 \\
\end{array}
\right . }$  \\ \hline

 $n = 10,000$ & $\displaystyle{\left . 
\begin{array}{cccc}  
-3.063005 & 1.445740 & 0.857643 & 0.759625 \\
0.725021 & 0.104546 & 1.146716 & -1.976280 \\
0.034128 & -0.078763 & 1.305917 & -1.261279 \\
0.549473 & -0.921383 & 0.058663 & 0.313245  \\ 
\end{array}
\right . }$  \\

\hline
\end{tabular}

\end{table}

%\skip

The matrices listed in Tables 3 and 4 indicate that for each $i, j = 1, \ldots, d$, $[\sqrt{n}(\widehat{P}_{n} - P)]_{ij}$ and $[\sqrt{n}(\widetilde{P}_{n} - P)]_{ij}$ remain bounded as $n \rightarrow \infty$ and that they are of the same order of magnitude. In fact, simulations up to $n = 1,000,000$ indicate exactly the same result. This example demonstrates by direct computation that $\widehat{P}_{n} - P = O(n^{-0.5})$ and $\widetilde{P}_{n} - P = O(n^{-0.5})$

\section{Simulation Study Structure}

The goal of this simulation study is to perform a quantitative comparison
between the
performance of the bootstrap method based on the maximum likelihood
estimator and its smoothed version. The true transition probability matrix
$P$ is known. To ensure that the structure of $P$ does not unduly 
influence the results, two different transition probability matrices were
used:
$$
P_{{\rm{I}}} = \left [ 
\begin{array}{ccc}
\frac{4}{10} & \frac{3}{10} & \frac{3}{10} \\
\frac{3}{10} & \frac{4}{10} & \frac{3}{10} \\
\frac{3}{10} & \frac{3}{10} & \frac{4}{10} \\
\end{array}
\right ]
{\textrm { and }} 
P_{{\rm{II}}} = \left [ 
\begin{array}{ccc}
\frac{2}{20} & \frac{9}{20} & \frac{9}{20} \\
\frac{9}{20} & \frac{2}{20} & \frac{9}{20} \\
\frac{9}{20} & \frac{9}{20} & \frac{2}{20} \\
\end{array}
\right ].
$$

For both of the true transition probability matrices, simulations
were
conducted for all the combinations of parameters $n = 25, 50, 100$ and $u
= 0.5, 1.0, 2.0$ and $\infty$. Note that $u = \infty$ corresponds to the
standard bootstrap.

Each simulation consists of the following steps:

\begin{enumerate}
\item A single chain of size $n$ is generated from the true transition probability matrix. Estimators $\widehat{P}$ and $\widetilde{P}$ are computed.
\item The bootstrap method (as described before) is
applied, using $\widehat{P}$ and $\widetilde{P}$, respectively. The number
of bootstrap resamples generated is $B = 5000$.
\item Bootstrap 90\% confidence intervals for the elements $P_{11}$ and
$P_{12}$ are computed, based on $\widehat{P}$ and $\widetilde{P}$,
respectively using the bootstrap percentile method outlined previously.
\item Steps 1-3 are repeated 1000 times and the observed coverage
properties of the intervals from the two estimators are compared.

\end{enumerate}

\section{Simulation Results and Conclusion}

\bigskip

\begin{table}
  \centering
  \caption{The empirical coverage of the standard ($u = \infty$) and
smoothed bootstrap percentile method confidence intervals for the entries
$P_{11}$ and $P_{12}$ of $P_{\rm {I}}$ and $P_{\rm {II}}$. The specified
nominal coverage is $90\%$.}\label{}   

\begin{tabular}{cccccc}
 &  & $P_{\rm {I}}$ &  & $P_{\rm {II}}$ & \\ \hline                  
$n$ & $u$ & $P_{11}$ & $P_{12}$ & $P_{11}$ & $P_{12}$ \\ \hline
25 & 0.5 & 90.6 & 90.6 & 99.6 & 99.6 \\
25 & 1 & 86.2 & 86.2 & 99.3 & 99.3 \\
25 & 2 & 81.6 & 81.6 & 53.0 & 53.0 \\
25 & $\infty$ & 81.5 & 85.4 & 53.0 & 85.6 \\
50 & 0.5 & 93.1 & 93.1 & 97.8 & 97.8 \\
50 & 1 & 86.8 & 86.8 & 79.3 & 79.3 \\
50 & 2 & 85.4 & 85.4 & 79.6 & 79.6 \\ 
50 & $\infty$ & 85.3 & 88.6 & 79.5 & 89.2 \\
100 & 0.5 & 92.0 & 92.9 & 94.2 & 94.2 \\
100 & 1 & 88.1 & 88.1 & 89.3 & 89.3 \\
100 & 2 & 87.1 & 87.1 & 82.7 & 82.7 \\ 
100 & $\infty$ & 87.0 & 89.1 & 82.4 & 90.2 \\ 
\hline
\end{tabular}

\end{table}

The results of the small simulation study are presented in Table 5 and 
seem to indicate that:

\begin{enumerate}
\item For almost all combinations of simulation parameters $n$ and $u$,
the coverage performance of the confidence intervals based on
$\widetilde{P}$ is better than for the intervals based on $\widehat{P}$.

\item At fixed chain length, $n$, increasing the smoothing parameter $u$
leads to narrower confidence intervals, with lower coverage performance.

\item Increasing the chain length leads to better coverage performance of
the standard confidence intervals. The effect this variation has on the
coverage performance of the smoothed intervals (at fixed $u$) is
inconclusive.

\item Overall, it appears that the best coverage performance (always
higher than the nominal value $90\%$) corresponds to the smallest value
allowed for the $u$, $u = 0.5$.

\end{enumerate}

%\section{References}

\end{document}